\RequirePackage{fix-cm}

\documentclass[twocolumn]{svjour3}         	

\smartqed  							

\usepackage{graphicx}
\usepackage{amssymb}
\usepackage[sort,numbers]{natbib}
\usepackage{overpic}
\usepackage{booktabs}
\usepackage{tabularx}
\usepackage{booktabs}
\usepackage{multirow, makecell}
\usepackage{graphicx}
\usepackage{amssymb, amsmath}
\usepackage{color}
\usepackage{xspace}
\usepackage{siunitx}
\usepackage{xspace}
\usepackage{siunitx}
\usepackage[export]{adjustbox}
\usepackage[usenames,dvipsnames,table]{xcolor}
\usepackage{rotating}
\usepackage{pdflscape}
\usepackage{lscape}
\usepackage{emptypage}
\usepackage{fancyhdr}

\newcommand{\psr}{PSR~B1259--63\xspace}

\def \psr {PSR~B1259--63}
\def \lsi {LS~I~+61~303}
\def \ls {LS~5039 }
\def \HESSJ0632 {HESSJ0632+057}
\def \1FGLJ1018{1FGL~J1018.6--5856}
\def \LMCP3{LMC--P3}
\def \PSR2032{PSR~J2032+4127}
\def \HESSJ1832 {HESSJ1832--093}

\def \J1023{PSR~J1023+0038}
\begin{document}

\begin{centering}
\title{\textbf{Phenomenology of} gamma-ray emitting binaries}
\end{centering}


\author{Josep Maria Paredes        \and
        Pol Bordas 
}

\authorrunning{Paredes \& Bordas} 

\institute{J.M. Paredes  and P. Bordas \at
              ICCUB, Dept. Quantum Physics and Astrophysics,\\
              Universitat de Barcelona. \\
              \email{jmparedes@ub.edu} \\         
               \email{pbordas@ub.edu}}  

\date{Received: date / Accepted: date}

\maketitle

\begin{abstract}
Gamma-ray emitting binaries (GREBs) are complex systems. Its study became in the last years a major endeavour for the
high-energy astrophysics community, both from an observational and a theoretical perspective. Whereas the accumulation
of observation time for most Galactic gamma-ray sources is typically leading to highly accurate descriptions of their steady
phenomenology, GREBs keep providing ?exceptions to the rule? either through long-term monitoring of known systems or
in the discovery of new sources of this class. Moreover, many GREBs have been identified as powerful radio, optical and
X-ray emitters, and may significantly contribute as well to the Galactic cosmic-ray sea. Their understanding implies, therefore,
solving a puzzle in a broad-band and multi-messenger context. In these proceedings we will summarise our current understanding
of GREBs, emphasising the most relevant observational results and reviewing a number of controversial properties.\footnote{This paper is the peer-reviewed version of a contribution selected among those presented at the Conference on Gamma-Ray Astrophysics with the AGILE Satellite held at Accademia Nazionale dei Lincei and Agenzia Spaziale Italiana, Rome on December 11-13, 2017.}.


\end{abstract}

\section{Gamma-ray $emitting$ binaries}
\label{intro}

Binary systems are an {established} class of high-energy (HE, 100~MeV$<E<$100~GeV) and very-high-energy (VHE, $E>100$~GeV) gamma-ray sources. In the last decade, a grand-total of about $\sim20$ systems have been detected either by ground-based Cherenkov telescopes or gamma-ray satellites. These gamma-ray $emitting$ binaries (GREBs) include a number of different binary-system sub-classes. GREBs whose spectral energy distribution (SED) peaks at $\gtrsim 1$~MeV are labelled \textit{gamma-ray binaries}. Systems powered by accretion onto a black hole or neutron star and displaying relativistic jets, {with SEDs peaking at keV X-ray energies}, are dubbed \textit{microquasars}. Thermonuclear bursts following strong accretion episodes onto the surface of a white dwarfs give rise to \textit{novae} explosions. Powerful stellar outflows developing strong shocks drive gamma-ray emission in \textit{colliding wind binaries}, and HE emission has also been claimed from recycled, non-accreting \textit{millisecond pulsars} in binaries. Despite this heterogeneous sample, they all share a common property: their emission physics can be constrained thanks to the periodic variation of the physical conditions taking place within and around the binary system.  \\
Below we briefly highlight a number of results recently reported for some of these GREBs, segregated by sub-system classes. We also emphasise some of the main unknowns in trying to interpret the origin of their gamma-ray emission. A list with the currently known GREBs is provided in Table~\ref{Table1}. The reader is referred to dedicated extended reviews on these sources for an accurate description on their phenomenology and theoretical interpretation (see, e.g. \cite{Mirabel2006, Paredes2011, Dubus2013}).  


\begin{table}[t]
\centering

\begin{tabular}{cccc}
\toprule
&  &  &  \\ 
\emph{ $\gamma$Bs}          &  \multicolumn{3}{c}{ \psr~\cite{Aharonian2005_PSRB1259}, \ls \cite{Aharonian2005_LS5039},  }   \\
				    &  \multicolumn{3}{c}{ \lsi~\cite{Albert2006}, HESS~J0632+057 \cite{Aharonian2007_HESSJ0632}, }   \\
				    &  \multicolumn{3}{c}{ 1FGL~J1018.6--5856 \cite{Corbet2011}, \LMCP3 \cite{Corbet2016},  }   \\
				    &  \multicolumn{3}{c}{  PSR~J2032+4127 \cite{Lyne2015}, HESS~J1832--093 \cite{HESS2015}  }   \\

&  &  &  \\

\emph{$\mu$}Qs 		    &  \multicolumn{3}{c}{ Cyg X-3 \cite{Tavani2009}, Cyg X-1 \cite{Albert2007}, SS433 \cite{Bordas2015}}   \\ 														
				    &  \multicolumn{3}{c}{ V404~Gyg \cite{Loh2016}, AGL~J2241+4454 ($^{*}$) \cite{Lucarelli2010}  }   \\

&  &  &  \\

\emph{CWB} &  \multicolumn{3}{c}{ Eta Carinae \cite{Tavani2009_EtaCar} }  \\

&  &  &  \\ 

\emph{novae} &  \multicolumn{3}{c}{ V407 Cyg 2010 \cite{Abdo2010_V407Cyg}, V1324 Sco 2012 \cite{Cheung2012ATel}, }   \\
		  &  \multicolumn{3}{c}{ V959 Mon 2012 \cite{Cheung2012_V959}, V339 Del 2013 \cite{Hays2013}, }   \\
		  &  \multicolumn{3}{c}{ V1369 Cen 2013 \cite{Cheung2013_NovaCen2013},  V5668 Sgr 2015 \cite{Cheung2015_NovaSagit2015}, }   \\
		  &  \multicolumn{3}{c}{ V5855 Sgr (= ASASSN-16ma) \cite{Li2016_ASASSN16ma}, }   \\
		  &  \multicolumn{3}{c}{ TCP J18102829--2729590 \cite{Li2016_TJ1810}, }   \\
		  &  \multicolumn{3}{c}{ Nova Lupus 2016 \cite{Cheung2016_NovaLupus2016}, }   \\
		  &  \multicolumn{3}{c}{ V679 Car \cite{Franckowiak2018}, V1535 Sco \cite{Franckowiak2018} }   \\

&  &  &  \\ 

\emph{MSPs} &  \multicolumn{3}{c}{ PSR~J1023+0038 \citep{Archibald2009},  XSS~J12270--4859 \citep{Martino2010}}   \\
		  &  \multicolumn{3}{c}{ PSR~B1957+20 \cite{Wu2012} }   \\

&  &  &  \\

\bottomrule
\end{tabular}
   \caption[mycaption]{List of identified GREBs known as of today. The first column denotes the type of system; the second column lists the actual sources belonging to each group. References are given to observational studies conducted with the last generation of $\gamma$ instruments that first pointed out the GREB nature of the source. 
    
($^{*}$) note that the microquasar nature of AGL~J2241+4454 is still to be confirmed.}
\label{Table1}
\end{table}


\section{Gamma-ray binaries}
\label{GRB}


Seven gamma-ray binaries ($\gamma$Bs) have been so far confirmed as sources of both HE and VHE $\gamma$-ray emission (see Table \ref{Table1}), with one additional candidate recently proposed: HESS J1832--093 (\cite{HESS2015a, Eger2016}; see also \cite{Mori2017}). Gamma-ray binaries are composed of a compact object and a non-degenerate companion star. The nature of the compact object is unconfirmed in all cases with the exception of \psr~ and PSR~J2032+4127, from which radio pulsations have been detected, pinpointing its pulsar origin. As for the companion star, two sub-groups are commonly proposed, depending on whether or not they feature a dense circumstellar disk. The first subgroup features O-type companion stars and displays a single-peak profile in their $\gamma$-ray light-curve, with the peak location along the orbit depending on the geometrical properties of the system. The second group features a O$e$ or B$e$ star, and displays several peaks in their light-curves. In some instances, these have been correlated with the times in which the compact object crosses the companion's circumstellar disk. From spectral grounds, $\gamma$Bs display differential fluxes $\propto E^{-\Gamma_{\gamma}}$ with averaged $\Gamma_{\gamma}$ {in the range 2.5 to 2.9}. No cutoff is apparent in their spectra, which extend up to energies of $\sim$tens of TeV.\footnote{Phase-resolved spectroscopy can provide more extreme values for $\Gamma_{\gamma}$, e.g. in \ls $\Gamma_{\gamma} = 1.8$ when the compact object is in its inferior conjunction, whereas $\Gamma_{\gamma}$ = 3.1 during superior conjunction. Note that an exponential cutoff power law model best fits \ls during its inferior conjunction.}
From a theoretical perspective, $\gamma$-ray emission from $\gamma$Bs harbouring a pulsar could be produced at the interface of the pulsar wind with that of the companion star. The emission would be produced in this case by particles accelerated at the shock interface, similar to the shock structures predicted for isolated pulsars (see e.g. \cite{Kennel1984}) but accounting for the much enhanced ram pressure of the companion's wind. Additionally, gamma-ray flares from pulsar-$\gamma$Bs could be driven by ``cold" electrons interacting with an external photon field \cite{Bogovalov2000, Khangulyan2012}. If $\gamma$Bs host instead a black hole and they are powered by accretion, gamma-rays could be produced along a yet undetected jet-like feature, resembling the behaviour observed in microquasars (see below).

\subsection{\textit{\textbf{$\gamma$Bs: open questions}}}

\begin{itemize}
%
\item \textit{$Powering~engine$}: Only in the case of the  $\gamma$B system PSR~B1259--63 and PSR~J2032+4127 the nature of the compact object, a neutron star, has been unambiguously identified. The debate is still open for other systems, in which it is still uncertain whether gamma-rays are produced either by accretion-driven jets  or by rotation-powered strong pulsar winds interacting with the nearby medium (see e.g. \cite{Dubus2006, Romero2007, Massi2013}).
\item \textit{$\gamma$-ray spectral components}: The presence of two separate components has been observed in the spectra of some $\gamma$Bs at energies above a few tens of GeV \cite{Hadasch2012}. An unambiguous interpretation for such double-component is still lacking, despite a number of scenarios having being proposed (see e.g. \cite{Zabalza2013} and references therein). Such a second component arising at VHEs is not apparent in all $\gamma$Bs, nor is detected in other GREBs.  
\item \textit{HE flares:} Three periastron passages of \psr~have been covered by current HE $\gamma$-ray satellites. A bright HE flare has been detected recursively, carrying a significant fraction of the pulsar spin-down power. Although several models have been proposed (see e.g. \cite{Kong2012, Khangulyan2012}), none of them can explain the flares consistently in a broadband MWL framework.
\item \textit{Light-curve profiles}: Light-curves in the few $\gamma$Bs known so far display in most cases distinct features which remain unexplained. These include asymmetric profiles in the light-curve of \psr; non-negligible fluxes at orbital phases where absorption should be severe in \ls; sharp dips and double-peak profiles in HESS~J0632+057; cycle-to-cycle variability of the main VHE peak in \lsi. Whether or not a unified picture can be applied to the whole  $\gamma$B class and account for these light-curve features needs to be investigated.


\end{itemize}

\section{Microquasars}
\label{MQ}


X-ray binaries displaying relativistic jets are dubbed microquasars ($\mu$Qs) in analogy with Active Galactic Nuclei (AGN) \cite{Mirabel1992}). Since AGNs are known sources of $\gamma$-rays, $\mu$Qs became since their discovery an obvious HE and VHE emitter candidate. Microquasars display distinct X-ray spectral states, thought to be the result of a variable accretion rate onto the compact object (either a neutron star or a stellar-mass black hole). Hard X-rays may be produced by persistent jets in the so-called $low/hard$ spectral state \cite{Markoff2001}, which could extend to higher, $\gamma$-ray energies. Moreover, non-thermal (synchrotron) emission from jet blobs has been resolved in the radio/IR band in several systems, implying the presence of highly energetic electrons that may also emit $\gamma$-rays through inverse Compton (IC; \citealp{Atoyan1999, Corbel2002}). In addition, at least in two $\mu$Qs the presence of baryons has been confirmed, through the detection of lines of highly ionised elements (in SS433 \cite{Migliari2002}, and in 4U~1630--47 \cite{DiazTrigo2013}). 

In the $\gamma$-ray domain, $\mu$Qs were claimed to be strong and variable $\gamma$-ray sources in the 80's, most notably in the case of Cyg X-3 (see a summary in Fig.~1 from \cite{Chardin1989}), although these detections resulted to be highly controversial. Cyg X-3 has been recently confirmed as a HE $\gamma$-ray source by AGILE and $Fermi$-LAT \cite{Tavani2009, LAT2009}. HE $\gamma$-ray emission from the $\mu$Q Cyg X-1 has also been recently reported \cite{Sabatini2013, Malyshev2013, Zanin2016}. Moreover, the analysis of six years of $Fermi$-LAT observations resulted in the detection of a $\gamma$-ray signal towards the $\mu$Q SS433 \cite{Bordas2015}. At VHEs, the MAGIC Collaboration reported a hint of detection from Cyg X-1 (at 4.1$\sigma$ statistical level, after trial-corrections \cite{Albert2007}). The search for VHE emission from other systems did not reveal so far any positive detection \cite{Saito2009, Archambault2013, HESS2018}. 

From a theoretical perspective, the production of $\gamma$-rays in $\mu$Qs has been studied in a number of scenarios: either invoking IC emission at the jet on the binary scales, where the photon field provided by the companion star is the strongest, or following hadronic interactions and $\pi^0$-decay, assuming that relativistic protons are present in the jets (see \cite{Romero2003, Dermer2006, BoschRamon2006} and references therein). In this hadronic context, $\mu$Qs have also been suggested to be significant contributors to the Galactic cosmic-ray sea \cite{Heinz2002}. Large-scale $\gamma$-ray emission at the jet/medium interaction regions has also been proposed \cite{Bosch2005, Bordas2009}, following again the AGN analogy.

\subsection{\textit{\textbf{$\mu$Qs: open questions}}}

\begin{itemize}
\item \textit{A small population}: $\gamma$-ray emission towards a few number of $\mu$Qs has been reported so far {(see Table~\ref{Table1})}. At TeVs, the {marginal} detection of Cyg X-1 by the MAGIC Collaboration needs to be confirmed at a higher statistical significance level. These detections amount therefore to just a few cases out of the tens of $\mu$Qs systems displaying a relatively large jet power and/or strong non-thermal activity at other wavelengths. It remains therefore to be understood the $\mu$Qs' limitation in producing detectable levels of $\gamma$-ray emission in a general case. A deeper knowledge of the physics behind state transitions can be crucial in this regard (i.e, if similar to the case of Cyg X-3). This may be particularly relevant for strong flaring episodes, as the one observed in GRS~1915+105 (\citealp{Mirabel1994}). Persistent emission, however, may also be expected (e.g., Cyg X-1 and SS433).
\item \textit{Jet physics}: understanding jet formation and propagation in $\mu$Qs can provide unique clues also for other fields/objects in high-energy astrophysics. Jet launching mechanisms have been postulated long ago (see e.g. \cite{Blandford1977, Blandford1982}). Still, many aspects keep unresolved: the conversion of accretion or black-hole rotation into powerful kinetic ejections, the jet composition, and the acceleration processes, are amongst the most relevant ones. The typically short time-scales related to $\gamma$-ray variability, and the periodic changes in the system and environments in $\mu$Qs, should be used to leverage to some extend some of these uncertainties.  
\item \textit{Contribution to Galactic cosmic-rays}: if $\mu$Q jets are in general baryon-loaded, as directly observed in SS433 and 4U~1630--47, they could contribute significantly to the Galactic cosmic-ray sea \citep{Heinz2002}. On the other hand, the association of the steady $\gamma$-ray flux towards SS433 may have only been possible given the extreme kinetic power of its jets, $\sim 10^{39}$~erg~s$^{-1}$. Using a similar efficiency in kinetic power to $\gamma$-ray flux any steady $\gamma$-ray emission and cosmic-ray production from any less powerful $\mu$Q would require much longer exposure times (for the same ambient conditions). A stacking analysis using the now accumulated $\sim 10$~yrs of GeV observations with the latest instruments could also be envisaged to constrain the cumulative contribution of $\mu$Q at these energies. 

\end{itemize}

\section{Classical novae}
\label{novae}

Classical novae (CNe) are a sub-class of cataclysmic variables, binary systems composed of a white dwarf accreting from a low-mass companion that has filled its Roche Lobe. Novae typically display bright optical flares produced by thermonuclear explosions on the surface of the white dwarf. In the last years, $\gamma$-ray emission has been (unexpectedly) detected from several CNe (see \cite{Franckowiak2018} and references therein). The first of such detections occurred in the symbiotic system V407 Cyg, distinguished by hosting a Mira giant secondary star featuring a dense stellar wind. $\gamma$-ray emission from CNe has been considered in a scenario in which these gamma-rays are emitted by particles accelerated at the shock between the nova ejecta and the companion's wind (see e.g. \cite{Tatischeff2007}). These models, however, were unsuccessful in explaining the detection of further CNe with the $Fermi$-LAT, as these systems are instead hosting main-sequence companion stars, providing therefore a much lower density circumstellar material. Alternatively, an IC origin for the $\gamma$-ray emission has also been proposed, e.g. for the case of V407~Cyg (see e.g. \citep{Martin2013}). 

As of today, 8 CN have been detected at HE $\gamma$-rays (\cite{Abdo2010, Ackermann2014, Cheung2016a, Cheung2016b, Li2016a, Li2016b}; see Table~\ref{Table1}) with two more candidates at a lower statistical significance level \citep{Franckowiak2018}.  At VHE, novae keep undetected \citep{Aliu2012, Ahnen2015}.
\subsection{\textit{\textbf{CNe: open questions}}}

\begin{itemize}
\item \textit{Emission mechanisms}: The origin of $\gamma$-ray emission from CNe has been studied in several scenarii. Gamma-rays could be the result of $\pi^0$-decay produced in the interactions of shock-accelerated protons with thermal protons. These shocks could be internal, that is, within the novae ejecta itself (see e.g. \cite{Metzger2014, Chomiuk2014}). A weak neutrino signal could also be expected in this case \cite{Metzger2016}. In the alternative IC-based models, internal shocks may also be taking place, but their contribution would be $\sim$negligible \cite{Martin2018}. 
\item \textit{VHE detection}: Gamma-ray emission from novae extending beyond $\sim 100$~GeV has not yet been detected. Although shock-accelerated particles could reach energies of several TeVs (see e.g. \cite{Tatischeff2007}), VHE fluxes may be too low for the current IACTs. The improved capabilities of future facilities (CTA) will further constrain the high-energy end of novae's gamma-ray spectra, potentially revealing these sources as a new class of VHE emitter.
%
\end{itemize}

\section{Colliding wind binaries}

\label{CWB}

Contrary to all other {GREBs}, colliding wind binaries (CWBs) do not harbour a compact object, but they are composed instead of two massive stars. Particle acceleration takes place in the shock interface of the two star winds, leading in turn to copious production of non-thermal emission that can eventually reach the HE/VHE domain (\cite{Benaglia2003}; see \cite{Becker2007} for a review). Observationally, only one of such systems has been detected in gamma-rays: Eta Carinae ($\eta$) \cite{Tavani2009_EtaCar, Abdo2010_EtaCar}; see also the recent report of a detection at VHEs \cite{Leser2017}). $\eta$-Car is however unique: it is composed of two extremely bright and powerful stars, a luminous and rare blue variable and an O or Wolf-Rayet star, and it is also distinguished by displaying bright emission in hard X-rays (\citealp{Leyder2008}). In the gamma-ray domain, $\eta$-Car's emission appears to be modulated by the orbital period ($\sim 5.5$~yrs). Flaring emission has been also claimed by AGILE \cite{Tavani2009_EtaCar}. Such flaring behaviour, however, has not yet been confirmed with the $Fermi$-LAT \cite{Abdo2010_EtaCar}.

From a theoretical perspective, the HE $\gamma$-ray emission from $\eta$-Car has been interpreted either as IC emission by electrons accelerated at the wind shock interface, or as the result of hadronic interactions and subsequent $\pi^0$ decay, where the dense winds serve as target for relativistic protons which are also accelerated in the wind-wind shock region \cite{Farnier2011}. On the other hand, $\gamma$-ray absorption at binary system length-scales can be severe in the system. HE $\gamma$-rays could also be the result from pair production and subsequent cascading in the intense soft X-ray photon field known to be present in in the source \cite{Reitberger2012}. Larger-scale emission may also be possible \cite{Ohm2010}, although this would not be able to explain the orbital modulation observed at HEs.

\subsection{\textit{\textbf{CWBs: open questions}}}

\begin{itemize}
\item \textit{Source~population}: Only $\eta$-Car stands as of today as the only CWB that has been detected at gamma-ray energies. Other systems with comparable energy budget and located relatively nearby have been studied (see e.g. \cite{Werner2013}), with no success. A reanalysis of the larger $Fermi$-LAT data set, making use of the recently delivered PASS 8 data, could significantly enhance the number of CWB detected (see e.g. \cite{Pshirkov2016})
\item \textit{Emission at VHEs }: The H.E.S.S. collaboration has recently announced the detection of $\eta$-car in the VHE domain \cite{Leser2017}. This will provide key information as it will constrain the cutoff known to be present in the source spectrum \cite{HESS2012EtaCar}. Theoretical models should be able to place a quantitative limit to the efficiency of shock acceleration processes and/or to constrain the properties of the stellar winds in this system.  
 
\end{itemize}


\section{Transitional millisecond pulsars and ``black widows"}
\label{psrs}


A few \textit{transitional} millisecond pulsars, switching from accretion to a radio pulsar stage, have been detected at HE  $\gamma$-rays: PSR~J1023+0038 \cite{Archibald2009} and XSS~J12270--4859 \cite{Martino2010}. HE $\gamma$-ray emission is also reported from the ``black widow" system PSR~B1957+20 \cite{Wu2012}. In addition to magnetospheric pulsed emission, PSR~B1957+20 displays a distinct component at $E \gtrsim 2.7$~GeV modulated with the system orbital period \citep{Wu2012}. This component has been suggested to arise from the intra-binary shock of star and pulsar winds. Alternatively it has been proposed that this component could be produced in an IC scenario from ``cold" pulsar wind electrons scattering off photons from the pulsar magnetosphere or coming from the companion star \cite{Wu2012}. At VHEs, no gamma-ray emission has been so far reported from any of these systems.

\subsection{\textit{\textbf{Transitional MSPs: open questions}}}

\begin{itemize}
\item \textit{Intra-shock scenario}: If HE $\gamma$-rays in these systems is produced in the shock between pulsar wind and the low-mass companion star, this would be reminiscent of the shock structure modelled in the case of $\gamma$Bs. Further investigation is needed, in particular making use of the non-thermal emission produced in this intra-binary shock at lower wavelengths (see e.g. \cite{Bogdanov2005}).
\item \textit{$\gamma$-rays from a ``cold" pulsar wind}: if the scenario proposed in \citep{Wu2012} is confirmed, this could have further consequences in the modelling of $\gamma$Bs, providing in particular insights into the unresolved mechanism responsible for the flaring episodes in PSR~B1259-63 \citep{Khangulyan2012}.
\item \textit{A link with $\mu$Qs and $\gamma$Bs?}: The system XSS~J12270--4859 displays a $\gamma$-ray/X-ray flux ratio $\sim0.8$ \citep{Martino2010}. This value is in between what is observed in the $\gamma$Bs ($\sim 6.2$--6.8) and in $\mu$Q Cyg X-3 ($\sim 0.01$--0.03). XSS~J12270--4859 could be therefore an intermediate case in between accretion- and rotation-powered GREBs. 
 
\end{itemize}

\section{Concluding remarks}
GREBs offer a unique opportunity to study particle acceleration and high-energy emission/absorption processes in a relatively well-constrained periodically changing environment. Still, differences in the nature of some of the sub-classes of binary systems discussed here could make it difficult to retrieve a common scenario able to explain the complex features observed in their light-curves as well as their spectral properties. Separate, in-depth studies of each of these systems seem to be more appropriate in this regard, which should also make use of the improved capabilities of new facilities being developed (e.g. CTA in the VHE domain) and the increasing available information provided by the monitoring of GREBs at different energy bands.

\label{remarks}

\bibliographystyle{aa_pol.bst} {\footnotesize\bibliography{proceedings}}



\end{document}